\documentclass[10pt,twocolumn]{article}

\usepackage{relsize}
\usepackage{multicol}
\usepackage{graphicx}
\usepackage{amsmath}
\usepackage{lipsum}
\usepackage{color}
\usepackage{siunitx}
\usepackage{authblk}
\usepackage{subfigure}
\usepackage[margin=2cm]{geometry}

\title{Dip-coating of suspensions}
\author[1,2]{Adrien Gans}
\author[2,3]{Emilie Dressaire}
\author[1]{B\'en\'edicte Colnet} 
\author[1]{Guillaume Saingier}
\author[4,5]{Martin Z. Bazant}
\author[1,3]{Alban Sauret}
\affil[1]{Surface du Verre et Interfaces, UMR 125 CNRS/Saint-Gobain, 93300 Aubervilliers, France}
\affil[2]{FAST, CNRS, Univ. Paris-Sud, Univ. Paris-Saclay, 91405 Orsay, France}
\affil[3]{Department of Mechanical Engineering, University of California, Santa Barbara, CA, USA (E-mail: asauret@ucsb.edu)}
\affil[4]{Department of Chemical Engineering, Massachusetts Institute of Technology, Cambridge, MA 02139, USA}
\affil[5]{Department of Mathematics, Massachusetts Institute of Technology, Cambridge, MA 02139, USA}
\date{ }
\begin{document}

\twocolumn[
    \begin{@twocolumnfalse}
        \maketitle
        \begin{abstract}
           Withdrawing a plate from a suspension leads to the entrainment of a coating layer of fluid and particles on the solid surface. In this article, we study the Landau-Levich problem in the case of a suspension of non-Brownian particles at moderate volume fraction $10\% < \phi < 41\%$. We observe different regimes depending on the withdrawal velocity $U$, the volume fraction of the suspension $\phi$, and the diameter of the particles $2\,a$. Our results exhibit three coating regimes. (i) At small enough capillary number $Ca$, no particles are entrained, and only a liquid film coats the plate. (ii) At large capillary number, we observe that the thickness of the entrained film of suspension is captured by the Landau-Levich law using the effective viscosity of the suspension $\eta(\phi)$. (iii) At intermediate capillary numbers, the situation becomes more complicated with a heterogeneous coating on the substrate. We rationalize our experimental findings by providing the domain of existence of these three regimes as a function of the fluid and particles properties. \\
            \medskip\medskip\medskip
        \end{abstract}
    \end{@twocolumnfalse}
]

\section{Introduction}

 The deposition of a liquid film on a surface due to the withdrawal of a substrate from a liquid bath is a standard process in everyday life and industrial applications.\cite{ruschak1985coating,scriven1988physics,quere1999fluid,rio2017withdrawing} Coating processes are now extensively used for industrial purposes: coating of wires, optical components, or fibers. In such situations, the control of the coating thickness is an important parameter. Current technological challenges require the development of new coatings with different fluids or substrates to improve materials performance. Because of these numerous applications, the dip coating process has been extensively studied in various geometries such as plates, fibers or tubes and with different fluids.\cite{quere1999fluid,ruckenstein2002scaling}

\medskip

 The prediction of the coating thickness for a plate withdrawn from a bath of Newtonian fluid has first been proposed in 1942 by Landau and Levich\cite{landau1942physicochim} and then Derjaguin.\cite{Deryagin} Their seminal work considered a solid surface withdrawn from a liquid reservoir of Newtonian fluid at velocity $U$, leading to the coating of the surface by a thin film. In this situation, the thickness of the liquid film depends on the withdrawal velocity $U$, the viscosity of the liquid $\eta$,  and the capillary length $\ell_c=\sqrt{\gamma/(\rho\,g)}$, where $\gamma$ and $\rho$ are the surface tension and the density of the fluid, respectively, and $g$ is the standard gravity. Indeed, the thickness of the coated film is set by a balance between the surface tension force, the viscous force, and gravity. The entrainment regime depends on the capillary number $Ca=\eta\,U/\gamma$, which describes the ratio of the viscous to the capillary effects. In the limit of small capillary number, typically $Ca<10^{-2}$, the thickness of the liquid film is described by balancing the viscous and surface tension forces only. Various studies have shown that, on a flat plate, and for a Newtonian fluid, the film thickness is then equal to $h \simeq 0.94\,\ell_c\,Ca^{2/3}$. At larger capillary number, when gravity dominates the surface tension effects, the film thickness becomes $h \sim \ell_c\,Ca^{1/2}$. Besides, in this regime, gravitational drainage occurs, which explains why the film does not have a constant thickness along the substrate. The entrainment of a liquid film of Newtonian fluid has been extensively studied to describe the flow properties\cite{Mayer:2012dt} as well as the shape of the static and dynamic menisci.\cite{maleki2011landau}

\medskip

The fluid dynamics associated with the withdrawal of a solid object from a liquid bath is a very active experimental and theoretical research topic, mainly because of the great diversity of fluids and substrates that can be used. Recent studies have highlighted how the thickness of the liquid film depends on several factors: the withdrawal speed, properties of the substrate, such as its roughness,\cite{krechetnikov2005experimental,seiwert2011coating} and properties of the fluid. For example, the dip coating in non-Newtonian fluids with different rheologies has been considered by various authors.\cite{tallmadge1970withdrawal,ro1995viscoelastic,afanasiev2007landau,maillard2014solid} Different studies have also examined the influence of surfactant on the film thickness and have shown that a thinning factor $\alpha$ must be included to take into account the modification induced by the presence of the surfactant.\cite{quere1997liquid,shen2002fiber,krechetnikov2005experimental,krechetnikov2006surfactant,delacotte2012plate} In summary, much of the previous work has considered the deposition of Newtonian or non-Newtonian homogeneous fluids. The situation becomes more complicated with suspensions, as solid particles are dispersed in the liquid phase.\cite{guazelli2011}

\medskip

Suspensions of particles are of primary interest as they are found in a large number of familiar products from food to personal care, as well as in industrial and geophysical materials. Besides, suspensions are involved in numerous industrial processes including the dispensing of liquid, the coating of surfaces \cite{Ross:1999br,Davard:2000ki,Iliopoulos:2005jd} or the cleaning of substrates.\cite{Fryer2009} Owing to the number of applications that involve the flow and transport of suspensions, the properties of those complex fluids have received considerable attention in the past.\cite{guazelli2011} Hence, suspensions and particle-laden flows have led to numerous studies to describe suspension rheology,\cite{leighton1987,zarraga2000characterization,boyer2011} inertial suspension flow in pipes,\cite{nott1994,lyon1998,morris1999,matas2004} shear-induced migration of particles,\cite{eckstein1977,butler1999,nott2011} imbibition of suspension,\cite{Holloway:2011di} or the clogging in confined flows.\cite{wyss2006,bacchin2011,agbangla2014,sauret2014,dressaire2017clogging} However, most of these studies have considered bulk flows of suspension, that are well described by constitutive rheological measurements.\cite{zarraga2000characterization,boyer2011} Yet, this approach cannot capture the complexity of the suspension entrainment observed during a dip coating process. In this situation, the thickness of the liquid film can become comparable to the particle size and the displacement of a particle will be controlled by the drag force exerted by the fluid and the capillary force due to the air/liquid interface deformation.\cite{Colosqui:2013ih,lubbers2014dense}

\medskip

Capillary effects are important when interfaces confine particles.\cite{kralchevsky2000capillary} Recent studies have shown the influence of particles on the detachment of droplets from a jet of suspension,\cite{furbank2004experimental,bonnoit2012,Miskin2012} which has important industrial applications, including printing. This canonical situation consists of a viscous fluid that is extruded from a nozzle and eventually generates a droplet. In a pure liquid, {\textit{i.e.}}, without particles, the filament becomes thinner until the pinch-off of the droplet owing to interfacial tension effects. Various studies have then considered the detachment of a drop of suspension and have shown that the effective viscosity can describe the early pinch-off dynamics. \cite{furbank2004experimental,bertrand2012dynamics} However, at later stages, when the thickness of the liquid thread is of the same order of magnitude as the particles, their presence becomes essential for the detachment of the drop.  Typically, the detachment becomes accelerated and depends on the volume fraction and size of the particles.\cite{furbank2004experimental,bonnoit2012,vanDeen2013} Even a small amount of particles, trapped in the liquid filament can modify the detachment dynamics of the droplet significantly. This unique situation cannot be explained with a classical Newtonian fluid description or conventional rheology of diluted suspensions as the presence of particles locally perturbs the thinning of the filament. 

\medskip

The combination of thin film flows and non-Brownian particles has received little attention, and previous work remains mainly qualitative. For instance, the experimental work of Buchanan \textit{et al.} reported patterns deposited from the runoff of bidisperse colloidal suspensions.\cite{Buchanan:2007fs} Understanding how non-Brownian suspensions flow in thin films is particularly important for dip coating processes when the liquid film entrained by the substrate has a thickness comparable to the particle diameter.\cite{Ghosh:2007ik} As a result, the capillary and drag forces exerted on the particles are comparable.\cite{Danov2000} During the dip coating process, the speed of the substrate withdrawal determines the thickness of the coated film, with faster withdrawals being associated with thicker coating films. When the coating fluid contains particles, the 2D numerical study of Colosqui \textit{et al.} shows that the withdrawal speed influences the entrainment of particles.\cite{Colosqui:2013ih} The numerical limitations only allowed to investigate a 2D situation with a limited number of particles. However, in practical applications, suspensions contain a significant amount of spherical particles, with a three-dimensional shape. The characterization of dip coating for a suspension made of non-Brownian and neutrally buoyant particles remains poor. Yet, understanding the regimes of dip coating of suspensions is critical to avoid its negative effects on coating applications, including local variations of the film thickness and introduction of asperities.

\medskip

In this paper, we consider the withdrawal of a smooth glass plate from a liquid bath of non-Brownian suspension. We measure the average thickness of the liquid film deposited on the plate using a gravimetry method. We characterize the three regimes encountered: (i) no particle entrainment at small capillary number leading to a coating of interstitial fluid alone, (ii) an heterogeneous state at moderate capillary number, in which the liquid film is thinner than the particle size and finally (iii) a coating layer of suspension, whose thickness can be described using the effective viscosity of the suspension. Here, we investigate the influence of the fluid properties and the particle size on the coating regime. We then rationalize our findings with the Landau-Levich law using the effective viscosity of the suspension to predict the film thickness. We also define the domain of existence of the three regimes.


\section{Experimental methods}

\subsection{Experimental setup}

Model suspensions of non-Brownian particles are prepared using different high density silicone oils : AP 100, AR 200 and AP 1000 (purchased from Sigma Aldrich) of respective dynamic viscosity $\eta_{AP100}=0.106\,{\rm Pa.s}$, $\eta_{AR200}=0.204 \,{\rm Pa.s}$ and $\eta_{AP1000}=1.088\,{\rm Pa.s}$ at $25^{\rm o}{\rm C}$, with respective density $\rho_{AP100}=1058\,{\rm kg.m^{-3}}$, $\rho_{AR200}=1042\,{\rm kg.m^{-3}}$ and $\rho_{AP1000}=1082\,{\rm kg.m^{-3}}$ and of surface tension $\gamma=21 \pm 1\,{\rm mN.m^{-1}}$. The particles are polystyrene (PS) particles of radius: $a=[20,\,40, \,70,\,125] \,{\rm \mu m}$ and average density of $\rho_p  = 1056 \pm 3\, {\rm kg.m^{-3}}$ (Dynoseeds from Microbeads). The PS particles are dispersed in the silicone oil AP 100 using a mechanical stirrer (Badger Air-Brush Paint Mixer) leading to a homogeneous dispersion of particles. The silicone oils used ensure that the particles are perfectly wetted by the liquid. The size of the particles is sufficiently large for both colloidal forces and Brownian motion to be negligible.\cite{zarraga2000characterization,boyer2011} The suspensions are neutrally buoyant over the timescale of our experiments. The volume fraction of particles, \textit{i.e.}, the ratio of the volume of particles $V_p$ to the total volume $V_{tot}=V_p+V_f$ (where $V_f$ is the volume of fluid), is denoted $\phi =V_p/V_{tot} \in [10,\,41]\,\%$. In each experiment, the suspension is prepared by mixing the particles and the liquid at the desired volume fraction $\phi$. We use pure silicone oils AR 200 and AP 1000 for comparison, as their viscosities approximately match the effective viscosity of the suspensions of volume fraction $\phi=18\%$ and $\phi=41\%$.

 \begin{figure}
 \includegraphics[width=0.5\textwidth]{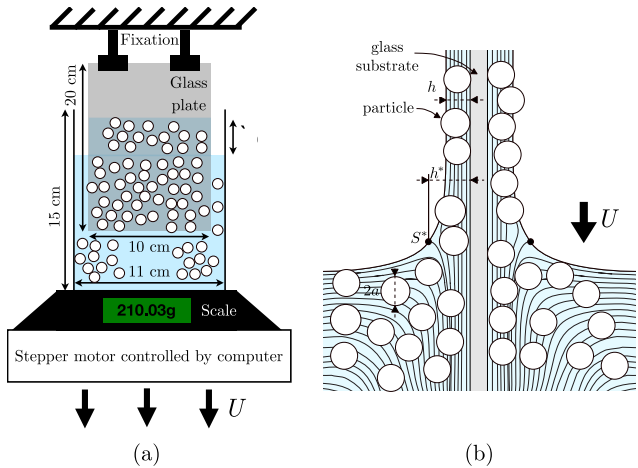}%
 \caption{Schematic of the dip coating experimental setup. (a) Front view showing the experimental setup and (b) side view of a zoom near the meniscus.\label{Fig_1}}
 \end{figure}

The experiments consist of withdrawing a glass plate (width: $w=10$ cm, height: $H=20$ cm, thickness : $e=2$ mm) from a container (width: 11 cm, height: 15 cm, thickness : 4 mm) filled with the suspension at a constant speed $U$ [Fig. \ref{Fig_1}(a)]. A high-precision stepper motor (Zaber) is controlled by computer and moves at speeds up to $2.4 \,{\rm cm \, s^{-1}}$ with a $\pm 2\,\%$ precision on the speed used. The glass plate is attached to a fixed support in the laboratory frame of reference while the stepper motor translates the container. The width of the glass plate is much larger than the thickness of the coated layer (smaller than 1 mm), validating the assumption of a 2D flow. The dip coating dynamics is recorded with a digital camera (Nikon D7100) and a 200 mm macro objective. Using silicone oil as the interstitial fluid, we did not notice any instabilities of the liquid film during the withdrawal of the glass plate from the liquid bath.

\subsection{Measurement of the average liquid thickness}

Because we cannot visualize through particles, we use the gravimetry method as previously successfully performed by various authors.\cite{krechetnikov2005experimental,Maleki:2011in,ouriemi2013experimental} The volume of suspension removed from the container during a withdrawal can be obtained from the difference of mass measured by a scale $\delta m$ and the fluid density $\rho$, with $V=\delta m/\rho$. Because of the presence of particles, the thickness of the suspension film can be non-uniform, and we report the average thickness deduced from the volume of fluid withdrawn and the coated area $A$ : $h=V/A$. The suspension is neutrally buoyant, $\textit{i.e.}$, the density of the particles and the fluid is the same, which allow using this method to determine the liquid thickness. Both sides of the plate are coated as illustrated in Fig. \ref{Fig_1}(b), which is taken into account in the total area.

  The main drawback of this method is the presence of a lower edge effect. Indeed, when the plate exits the liquid bath, a film is created between the bottom of the plate and the liquid bath, which will eventually break. This effect induces an additional mass on the lower edge of the plate and compromises the accuracy of the measurements. Thus, similarly to Ouriemi \textit{et al.},\cite{ouriemi2013experimental} we performed two withdrawal experiments with different dipping lengths $l_2>l_1$. Subtracting the masses obtained with the two dipping lengths cancels out the volume of fluid that remains attached to the bottom of the plate after the withdrawal. The average thickness on the plate $ h $ is then determined by the expression
\begin{equation}
h=\frac{1}{(l_2-l_1)\,\rho\,p}\left(\frac{\delta m_2}{N_2}-\frac{\delta m_1}{N_1}\right)
\end{equation}
where $l_2>l_1$, $m_1$ and $m_2$ are the masses measured for the two dipping lengths $l_1$ and $l_2$, respectively. $N_1$ and $N_2$ are the numbers of withdrawals and $p=2\,(w+e)$ is the wetted perimeter.

\subsection{Dip coating of Newtonian liquids}

 \begin{figure}
 \includegraphics[width=0.5\textwidth]{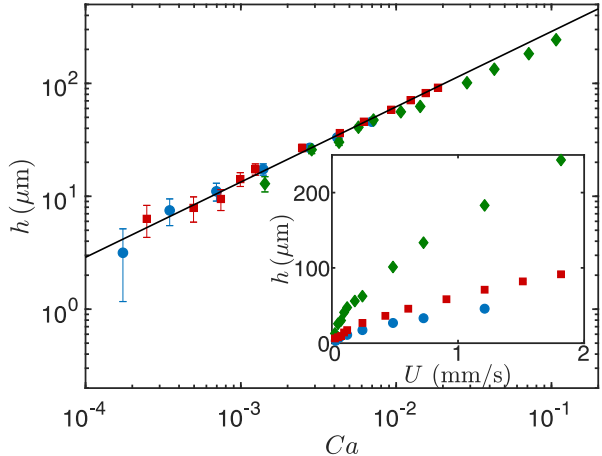}%
 \caption{Average thickness $h$ entrained on the glass plate withdrawn from the bath of silicone oil as a function of the capillary number $Ca$. The dotted line is the Landau-Levich-Dejarguin (LLD) law (\ref{LLDLaw}). Inset: non-rescaled data showing the thickness $h$ of the liquid film when increasing the withdrawal velocity $U$. In both figures, the symbols indicate the viscosity of the silicone oil: $\eta_{AP100}$ (blue circles), $\eta_{AR200}$ (red squares) and $\eta_{AP1000}$ (green diamonds).\label{Fig_2}}
 \end{figure}

\begin{figure*}
 \begin{center}
 \includegraphics[width=0.9\textwidth]{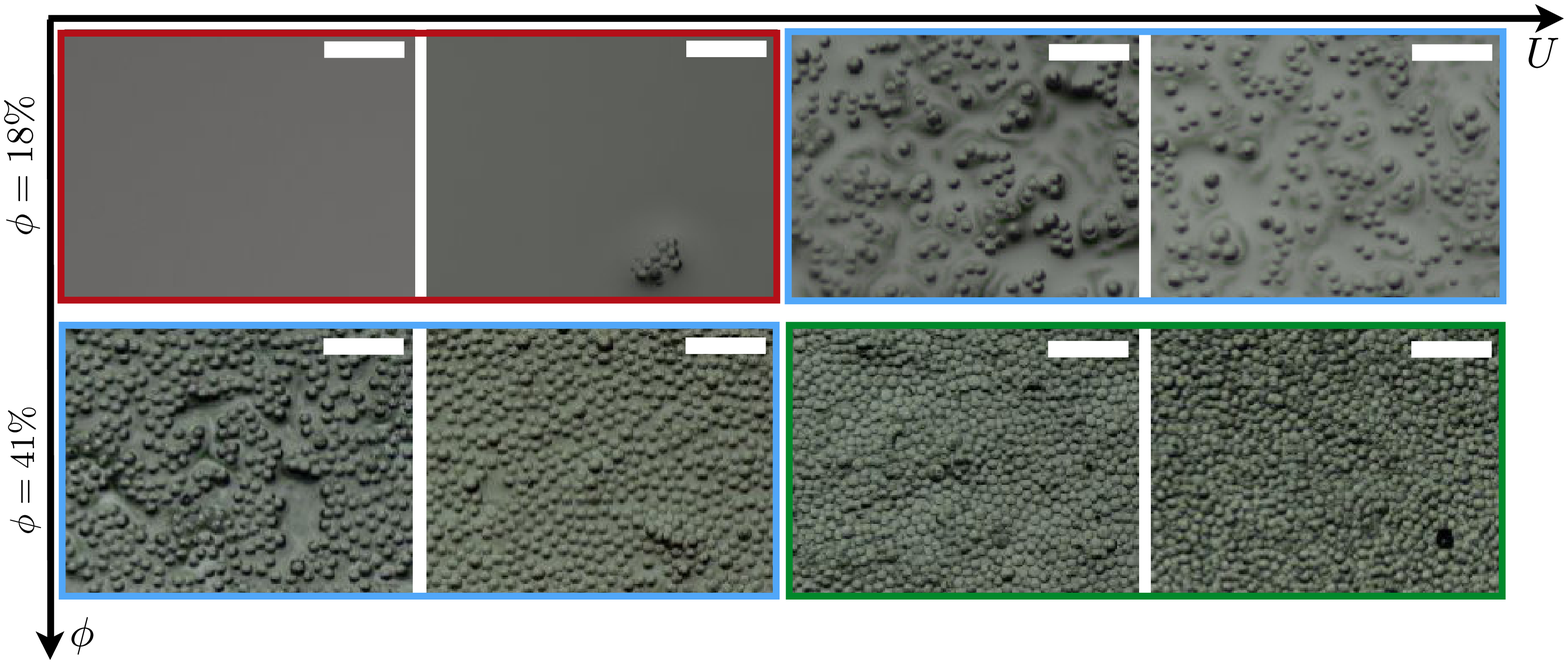}%
 \caption{Pictures taken after the withdrawal of the plate from suspensions of concentration, $\phi=18\%$ and $\phi=41\%$ of PS particles of diameter $2\,a=140\,\mu{\rm m}$ dispersed in the Silicone oil AP 100. From left to right, the withdrawal velocities are $U=24,\,121,\,484,\,1453\,\mu{\rm m.s^{-1}}$. Scale bar is 1000 $\mu$m and the color box indicates the different regimes: liquid only (red), heterogeneous coating (blue) and multilayer of particles (green).\label{Fig_3}}
 \end{center}
 \end{figure*}

We first validate the experimental method by performing the classic dip coating experiments with particle-free silicone oil. We measure the thickness of the liquid film, and we determine the stability of the system.

We report the experimental data in the inset of Fig. \ref{Fig_2} for the three silicone oils considered in this study: AP 100, AR 200, and AP 1000. As expected, the thickness of the film entrained on the plate increases with the withdrawal speed and viscosity. When a plate is withdrawn from a bath of Newtonian fluid of viscosity $\eta$ and surface tension $\gamma$, the thickness of liquid deposited is a power-law of the capillary number $Ca= \eta \, U / \gamma $. In the viscous-capillary regime where  the effects of gravity are negligible, the thickness of the liquid layer $h$ is uniform along the plate and is given by the theoretical prediction of Landau and Levich:\cite{landau1942physicochim,Maleki:2011in}
\begin{equation}\label{LLDLaw}
h=0.94\,\ell_c\,Ca^{2/3}
\end{equation}
where $ \ell_c = \sqrt{\gamma/(\rho\,g)}$ is the capillary length of the fluid. We rescale our data accordingly for the three silicone oils in Fig. \ref{Fig_2}. The typical uncertainties on the measurements in the experiments are represented in this figure. The uncertainties depend on the mass of liquid entrained by the plate, thus on the liquid film thickness only. In the following, we only plot the typical uncertainties for one set of experiments. Here, the data for the three Newtonian fluids without particles are in good agreement with the Landau-Levich-Dejarguin (LLD) law in the range $Ca\in [10^{-4},2\,\times 10^{-2}]$. In all situations, the liquid film is thick enough for long-range forces to be negligible. At large capillary number, typically more than $10^{-2}$, the effects of gravity become important leading to a viscous-gravity regime and a deviation from the LLD law. We now consider the dip coating of a suspension of non-Brownian and neutrally buoyant particles in the silicone oil at a volume fraction $\phi$.

 \section{Results: three coating regimes}

\subsection{Qualitative observations}

We begin by characterizing the macroscopic structure of the coating layer. Previous work has mostly been devoted to colloidal suspensions, whose liquid phase evaporates during the coating.\cite{Ghosh:2007ik} In this situation, experiments highlight the presence of periodic horizontal assemblies of particles. The width and spacing of the stripes depend on the withdrawal velocity $U$. More recently, diluted suspensions of particles have been considered through numerical simulations.\cite{Colosqui:2013ih} The particles are entrained in the liquid film forming clusters within the meniscus or later in the liquid film. However, because of the numerical cost, only a few particles have been considered in this study. 

Here, we perform experiments using suspensions with a volume fraction of $\phi \in [10,\,41]\,\%$. No evaporation occurs over the time scale of the experiments. We observe qualitatively different coating regimes, as illustrated in Fig. \ref{Fig_3}.

From a suspension of particles of diameter $ 2 \, a = 140 \, \mu {\rm m}$ and a volume fraction $ \phi = 18 \% $, the withdrawal of a plate at low velocity $U$ over a length of $60\,{\rm mm}$ results in a coating film without particles. The thickness of the liquid film is not sufficient to entrain the particles, which remain trapped in the meniscus. When the withdrawal speed $U$ increases, a few clusters of particles may be observed on the glass plate but they do not form a regular arrangement, and most of the plate remains coated by liquid alone. This first regime is referred to as \textit{liquid only}.

   Beyond a first threshold velocity, more particles are entrained in the liquid film forming a regular pattern. The particles tend to form clusters on the plate because of the capillary attraction between them.\cite{vella2005cheerios,cavallaro2011curvature} The volume fraction of the suspension being moderate here, the particle distribution is heterogeneous with regions where clusters of particles are present and others only covered by a liquid film. In this regime, the number of particles entrained per surface area increases as the withdrawal speed increases. For practical applications, such coating is likely undesirable as it results in a surface with heterogeneous properties. We will refer to this regime as the \textit{heterogeneous coating regime}. Varying the volume fraction of the suspension (here $\phi=41\%$), we observe that the particle concentration influences the threshold velocity between the liquid only regime and the heterogeneous coating regime. 

   When the withdrawal velocity $U$ further increases, a second threshold value is reached, and a uniform coating of particles and fluid is obtained. This coating can consist of multiple layers of particles depending on the volume fraction of the suspension $\phi$. This regime is referred to as the \textit{effective viscosity regime}.

In summary, as illustrated in Fig. \ref{Fig_3}, we highlight three regimes, whose existence depends on the withdrawal velocity $U$, the viscosity of the interstitial fluid $\eta_0$, the size of the particles $2\,a$ and the volume fraction $\phi$ considered. First, (i) at low withdrawal velocity $U$, only the liquid is entrained on the plate, and the particles remain trapped in the bath of suspension. This leads to a uniform coating by the interstitial fluid only. Then, (ii) at a higher velocity, a heterogeneous coating is observed with clusters of particles randomly distributed on the plate. In this regime, the number of particles per unit area increases with the withdrawal velocity. Finally, (iii) at large withdrawal velocity a uniform coating of particles, which can be composed of several layers is obtained. To characterize the particle-laden film thickness corresponding to these different regimes, we rely on macroscopic measurements. Indeed, it is challenging to get a local measurement of the film thickness when both particles and fluid form a heterogeneous coating because of the deformation of the air/liquid interface and the presence of particles. In the following, we use the gravimetry method presented in the previous section to characterize the average thickness of the deposited suspension film and identify the domain of existence of the different coating regimes.

\subsection{Average thickness of the liquid film}

  \begin{figure}
  \includegraphics[width=0.5\textwidth]{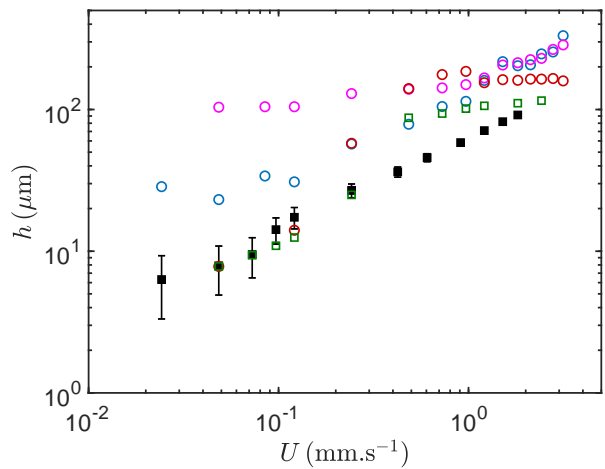}%
  \caption{Examples of average thickness $h$ of the coating layer after the withdrawal of the plate when increasing the withdrawal speed $U$ for the silicone oil AR 200 (filled black squares) and different suspensions of PS particles in silicone oil AP 100: diameter $2\,a=20\,\mu{\rm m}$ and a volume fraction $\phi=41\%$ (blue circles), $2\,a=140\,\mu{\rm m}$ and $\phi=41\%$ (magenta circles), $2\,a=140\,\mu{\rm m}$ and $\phi=18\%$ (green circles) and $2\,a=250\,\mu{\rm m}$, $\phi=18\%$ (red circles). The typical uncertainties on the thickness are shown for the silicone oil AR 200 (black squares) and are of the same order of magnitude for all data sets.\label{Figure4_RawSuspension}}
  \end{figure}

We measure the average thickness $h$ of the coated layer after the withdrawal of the plate in Fig. \ref{Figure4_RawSuspension}.  Whereas for the pure silicone oil, the average thickness increases in a monotonic way with an exponent of about $h\propto U^{2/3}$, the situation is more complex in presence of particles. In the heterogeneous regime, we should emphasize that the coating film is not flat. However, the average value of the thickness is sufficient to define the domain of existence of the three coating regimes.

 Here, the evolution of $h$ for the thin film of suspension does not follow the monotonic trend reported for the pure silicone oil in Fig. \ref{Figure4_RawSuspension}. The evolution of the coating thickness $h$ with the withdrawal velocity $U$ depends both on the size of the particles $2\,a$ and the volume fraction of the suspension $\phi$. For a given particle size, the average thickness of the liquid film $h$ can remain nearly constant over a range of withdrawal velocity, typically around $h \sim 2\,a$. For example, a constant average thickness of $h \sim 20-30\,\mu{\rm m}$ is observed between 0.02 and $0.1\,{\rm mm.s^{-1}}$ for particles of $20\,\mu{\rm m}$ and $h \sim 100-150\,\mu{\rm m}$  between 0.05 and $1\,{\rm mm.s^{-1}}$ for the $140\,\mu{\rm m}$ particles at $\phi=41\%$. 

 A sudden increase of the average thickness can also be observed in our measurements, for instance at $U \sim 0.6\,{\rm mm.s^{-1}}$ for the $140\, \mu{\rm m}$ particles at $\phi=41\%$. This sudden variation corresponds to the transition between two coating regimes, for instance between the regime where only the liquid is coating the plate to the entrainment of particles in the heterogeneous regime. These first results suggest that the regimes previously identified depend on the withdrawal velocity $U$, the particle diameter $ 2\, a $ and the volume fraction of the suspension $\phi$, and lead to different entrainment mechanisms.

 \subsection{Capillary number of the suspension}

 \begin{figure}
\includegraphics[width=0.5\textwidth]{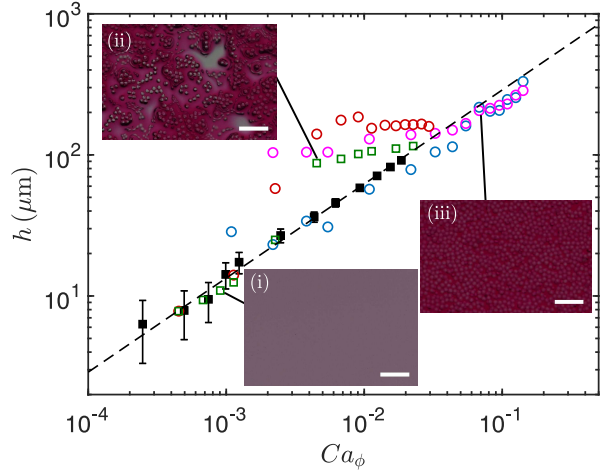}
 \caption{Examples of thickness of the coated layer $h$ as a function of the capillary number based on the effective viscosity of the suspension $Ca_\phi$. The color code is the same as in Fig. \ref{Figure4_RawSuspension}. The solid line shows the LLD law $h=0.94\,\ell_c\,{{Ca_\phi}^{2/3}}$. The insets show typical coating patterns observed on the withdrawn plate for $2\,a=140\,\mu{\rm m}$ particles at (i)-(ii) $\phi=18\%$ and (iii) $\phi=41\%$ for different values of ${Ca_\phi}$. Scale bars are 1000 $\mu$m. The typical uncertainties on the thickness are shown for the silicone oil AR 200 (black squares) and are of the same order of magnitude for all data sets.\label{Fig_5a}}
 \end{figure}

Since our experiments highlight that the average thickness of the film $h$ depends on the volume fraction $\phi$, we consider the suspension as a continuum fluid having an effective viscosity $ \eta(\phi)$. We can then define a capillary number based on the viscosity of the suspension $Ca_\phi = \eta(\phi) \, U / \gamma$. As evidenced by Bonnoit \textit{et al.}, the Zarraga model describes well the shear viscosity of a suspension of non-Brownian and neutrally buoyant polystyrene particles in the range of volume fraction considered here ($\phi \in [5\%,\,41\%]$).\cite{bonnoit2012} We then estimate the viscosity of the suspension using the relation:\cite{zarraga2000characterization}
\begin{equation}
\eta(\phi)=\eta_{0}\,\frac{\text{exp}\left(-2.34\,\phi\right)}{\left(1-\phi/\phi_m\right)^3},
\end{equation} 
where $\eta_0$ is the viscosity of the interstitial fluid, here the viscosity of the silicone oil AP 100, and $\phi_m= 0.62$ denotes the maximum packing fraction. The values of the shear viscosity obtain using this relation capture well our experimental measurements of the viscosity. The modified capillary number based on the effective viscosity of the suspension becomes
\begin{equation}
Ca_\phi = \frac{\eta_0 \, U}{\gamma}\,\,\frac{\text{exp}\left(-2.34\,\phi\right)}{\left(1-\phi/\phi_m\right)^3}.
\end{equation}

In Fig. \ref{Fig_5a}, we report experimental results rescaled with $Ca_\phi$. For all suspensions considered, we observe that the average thickness $h$ follows the LLD law using the effective viscosity of the suspension.  This corresponds to the effective viscosity regime (iii) for large enough capillary number, \textit{i.e.}, large enough withdrawal velocity.

 The size of the particles $2\,a$ is found to influence when the ``effective LLD law'' is recovered: for small particles, the \textit{effective viscosity} regime is obtained at a smaller capillary number. Typically, the LLD regime is observed as soon as the thickness of the coating films becomes equivalent to the size of the particles, $h \sim 2\,a$. We also note that the drainage of the suspension is limited compared to that of a Newtonian fluid having an equivalent viscosity. Indeed, the particles deform the liquid interface, which generates an interfacial force keeping the particles on the plate.

At small capillary number, in the \textit{liquid only} regime, our experiments show that no particles are entrained in the coating film [Fig. \ref{Fig_3}]. The particles remain trapped in the bath of suspension. At such low capillary number, corresponding to small withdrawal velocity, the thickness of the liquid film is much smaller than the particle size, typically smaller than few tens of micrometers and no particles are entrained. In this situation, the average thickness of the coated layer follows the LLD scaling law.

Finally, between these two regimes, the situation is more complicated with particles heterogeneously deposited on the plate. The liquid film between the particles is thinner than their diameter. In this (ii) \textit{heterogeneous} regime the average thickness $h$ remains roughly constant or increases only slightly over the intermediate range of capillary number increasing sharply at the onset of the (iii) \textit{effective viscosity} regime.

 \subsection{Evolution of the surface density of particles}

\begin{figure}
  \includegraphics[width=0.5\textwidth]{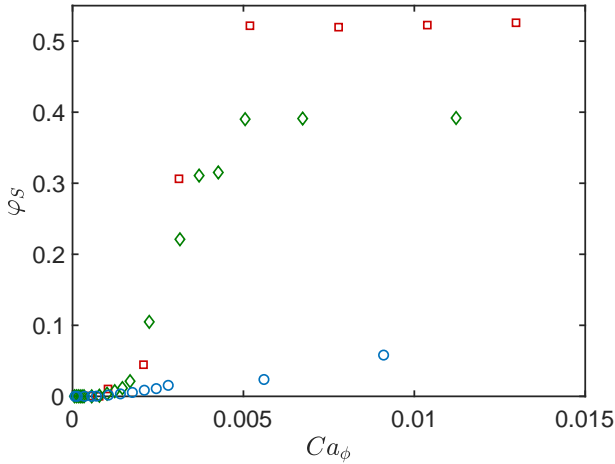}%
  \caption{Example of the evolution of the surface coverage $\varphi_S$ on the plate with the capillary number of the suspension $Ca_\phi$ for $140\,\mu{\rm m}$ particles at $\phi=5\%$ (blue circles), $\phi=18\%$ (green diamonds) and $\phi=41\%$ (red squares) dispersed in the AP 100 silicone oil.\label{surfacearea}}
  \end{figure}

In addition to the average thickness of the liquid film, we measure the surface density of particles entrained on the plate when increasing the capillary number $Ca_\phi$. The surface density of particles $\varphi_S = S_{part} / S_{tot} $ is defined as the ratio of the particle covered area to the total area of the plate. In Fig. \ref{surfacearea}, we report an example of this measurement for a particle diameter of $140\,\mu{\rm m}$. The data show that no particle is entrained on the plate at low withdrawal velocity. This part of the figure corresponds to the regime where only a liquid film is entrained on the plate. Then, increasing the capillary number $Ca_{\phi}$ leads to an increase of the surface density of particles $\varphi_s$. The value of the surface density depends on the volume fraction of the suspension. Beyond a given capillary number the surface density of particles saturates at a given value, which also depends on the volume fraction $\phi$. A theoretical description of this evolution, which occurs in the heterogeneous regime, is beyond the scope of the present paper. Nevertheless, these results show that even if the average thickness of the coating film remains almost constant over the intermediate range of capillary number in the regime (ii), the composition of the entrained film varies with the capillary number.

 \section{Discussion: transition between the different regimes}

 The domains of existence of the different regimes depend on the particle size $a$ and the concentration of the suspension $\phi$. To estimate the order of magnitude of the capillary numbers for which each regime is observed, we consider both the transition between (1) the liquid film only (i) and the heterogeneous coating regime (ii) and (2) the heterogeneous coating regime (ii) and the effective viscosity regime (iii). Each transition is characterized by a threshold capillary number.

\subsection{From (i) liquid films to (ii) heterogeneous films}

\begin{figure}
 \includegraphics[width=0.5\textwidth]{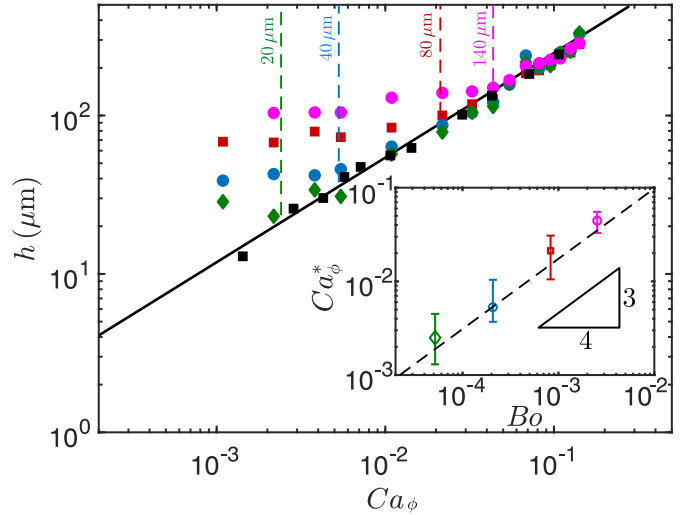}%
 \caption{Average thickness of the coating layer as a function of the capillary number $Ca_\phi$ for varying size of particles (green diamonds: 20 $\mu$m, blue circles: 40 $\mu$m, red squares: 80 $\mu$m, magenta circles: 140 $\mu$m) and a volume fraction $\phi=41.6\%$ and the silicone oil AP 1000 (black squares). The typical uncertainties on the experimental thickness are the same than in Fig. \ref{Fig_5a}. The solid line shows the LLD law $h=0.94\,\ell_c\,{Ca_\phi}^{2/3}$. The vertical colored dashed lines show the transition between the intermediate and effective viscosity regimes for the different diameters of particles considered. Inset: capillary number at which the suspension starts to follow the LLD law when using the effective viscosity of the suspension $Ca_{\rm \phi}^*$ as a function of the Bond number $Bo=(a/\ell_c)^2$. The dash-dotted line is the theoretical equation $Ca=3.1\,Bo^{3/4}$.\label{Fig_4}}
 \end{figure}

To determine transition (1) between the liquid film only and the heterogeneous coating regime, we calculate the condition of entrainment of a particle in the liquid film. We should emphasize that the presence of multiple particles, \textit{i.e.}, clusters will provide an additional force and will thus decrease the corresponding threshold capillary number by increasing the drag force. Nevertheless, this approach provides a first estimate of the threshold capillary number. At very small capillary number, no particles are entrained in the liquid film. We thus assume that the shape of the meniscus can be calculated from a pure liquid, \textit{i.e.}, the situation without particles.
Following Colosqui \textit{et al.}, we assume that a particle can be entrained in a liquid film if the thickness of the film at the stagnation point [$S^*$ in Fig. \ref{Fig_1}(a)] is larger than the particle diameter \cite{Colosqui:2013ih}. At this point, the liquid film has a thickness $h^*$, which leads to the following entrainment condition $h^*> 2 \, a.$ The stagnation point is the position where the speed of the interface vanishes and is related to the thickness of the liquid film away from the meniscus through the relation $ h^* = 3 \, h $.\cite{krechetnikov2010application} Hence, particle entrainment corresponds to $ h> 2 \, a /  3$. Using the theoretical thickness of the pure liquid film $h=0.94\,\ell_c\,Ca^{2/3}$, and defining the Bond number as $Bo=\left({a}/{\ell_c}\right)^2$, the existence of the liquid only regime corresponds to
\begin{equation}
Ca<0.6\,Bo^{3/4}.
\end{equation}
This condition corresponds to the critical condition for which the particles can assemble inside the meniscus, suggested in Ref. \cite{Colosqui:2013ih}.

For the particles of diameter 140 $\mu$m and the silicone oil AP 100, this condition corresponds to capillary numbers smaller than $ 6 \times 10^{- 3} $. Experimentally, the transition is observed at $Ca \sim 10^{-3}$ in Fig. \ref{Figure4_RawSuspension}. Indeed, for $Ca$ smaller than the predicted value, the particles accumulate in the meniscus and assemble into clusters. As a result, the viscous drag force exerted on the cluster becomes sufficient to overcome the capillary force that keeps particles in the liquid bath at low capillary number $Ca$. Thus, the entrainment of clusters can be observed at smaller capillary number than predicted. Our estimate for single particle entrainment provides an upper bound for the domain of existence of the first regime, the liquid only regime but the transition remains complex to derive analytically, because of the complexity of the cluster assembly. We should also emphasize that because of the large size of the particles, no clusters are formed within the bulk. Depending on the withdrawal velocity, clusters can form either in the meniscus or in the liquid film because of capillary attraction between particles. More specifically, Colosqui \textit{et al.} showed numerically in 2D that, for $Ca<0.6\,Bo^{3/4}$, the clusters of particle assemble in the meniscus, whereas for $Ca>0.6\,Bo^{3/4}$, the particles are expected to assemble in the liquid film.\cite{Colosqui:2013ih}

\begin{figure*}
 \includegraphics[width=1\textwidth]{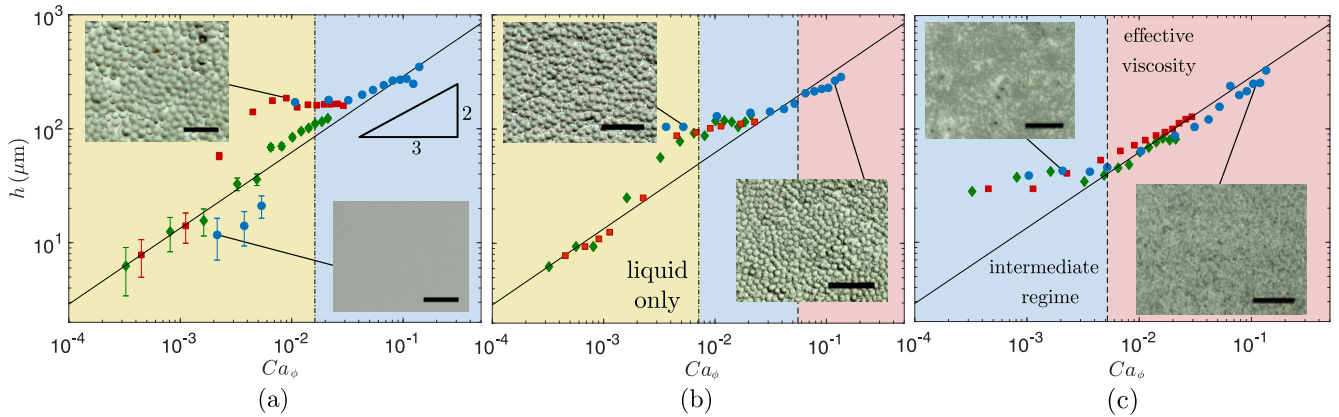}%
 \caption{Average thickness of the coating layer $h$ as a function of the capillary number $Ca_\phi$ based on the effective viscosity of the suspension for varying size of particles dispersed in silicone oil AP 100 (a) $250\, \mu{\rm m}$, (b) $140 \,\mu {\rm m}$ and (c) $40 \,\mu{\rm m}$ for a volume fraction $\phi=10\%$ (green diamonds), $\phi=18\%$ (red squares), and $\phi=41\%$ (blue circles). The typical uncertainties on the thickness are shown in (a). The solid line shows the LLD law $h=0.94\,\ell_c\,Ca^{2/3}$, the vertical dashed line and dash-dotted line show the theoretical prediction transition between the intermediate and effective viscosity regimes and between the liquid only regime and the intermediate regime, respectively for the different size of particles. The insets are pictures of the coated layer for a volume fraction of $\phi=41\%$. Scale bars are $1000\, \mu{\rm m}$.\label{Panel}}
 \end{figure*}

\subsection{From the (ii) heterogeneous films to the (iii) the effective viscosity regime}

We now consider transition (2) between the heterogeneous regime and the effective viscosity regime. For sufficiently diluted suspension, we assume that the suspension film behaves like an equivalent fluid with an approximate homogeneous coating thickness. As shown previously, the thickness of the liquid film is well defined by the LLD law using the effective viscosity of the suspension. The effective viscosity regime, or bulk description, can be assumed to apply when the coating film has a thickness $h$ greater than the particle diameter $2\,a$. This condition is equivalent to $ h> 2 \, a $, where $h$ is obtained using the capillary number based on the effective viscosity $Ca_\phi$. We therefore have
\begin{equation}\label{condition_size}
Ca_\phi>3.1\,Bo^{3/4}.
\end{equation}

In this approach, we have neglected that particles can deform the interface in the meniscus when the volume fraction of the suspension is large. We therefore expect that the threshold capillary number will be larger as more particles accumulate in the meniscus. Indeed, Bonnoit \textit{et al.} have studied the flow of a dense suspension on an incline, and report a deviation from the Newtonian viscous behavior at small film thickness.\cite{bonnoit2010inclined} More specifically, below a particular layer thickness, the viscosity is no longer constant, and the measured viscosity decreases. The cross-over thickness depends on the concentration of the suspension and the size of the particles. Therefore, in some situations, we expect the threshold capillary number to be slightly larger than our theoretical prediction. However, since we consider semi-dilute suspensions with a maximum volume fraction of $\phi=41\%$, our estimate should capture reasonably well the experimental results. In the following paragraphs, we compare the threshold condition in Eq. (\ref{condition_size}) with the experimental results obtained varying the properties of the suspension.

\subsection{Influence of the size of particles $2\,a$} 
 
We first characterize the influence of the particle size on the average film thickness $h$ on the plate. The experimental results are shown in Fig. \ref{Fig_4}, using the capillary number based on the effective viscosity of the suspension $Ca_\phi$. The solid line indicates the LLD law $h=0.94\,\ell_c\,{Ca_\phi}^{2/3}$.

The reported trends indicate that the transition between the heterogeneous film and the effective viscosity regime depends on the diameter of the particles, for a fixed volume fraction, $\phi=41.6\%$, as expected according to Eq. (\ref{condition_size}). The effective viscosity regime is observed for a capillary number slightly larger than  $Ca_\phi \sim 10^{-3}$ for particles of diameter $ 20 \, \mu {\rm m} $, which is in fairly good agreement with the predicted value. In the inset of Fig. \ref{Fig_4}, we report the measured value of the threshold capillary number $Ca_{\phi}^*$ for the sizes of particles considered as well as our theoretical prediction of the threshold, $Ca^*_\phi \simeq 3.1\,Bo^{3/4}$. The experimental results compare well with our model over the range of Bond numbers considered in this study. Below $Ca^*$, the coating film is heterogeneous, and the effective viscosity of the suspension is not sufficient to describe the entrained layer.

\subsection{Influence of the volume fraction of the suspension $\phi$}

We then consider the influence of the volume fraction, as illustrated in Fig. \ref{Panel}(a)-(c) for three sizes of particles, $250\,\mu{\rm m}$, $140\,\mu{\rm m}$ and $40\,\mu{\rm m}$. We observe that the transition to the effective viscosity regime does not seem to depend significantly on the volume fraction $\phi$ of the suspension in the range considered in this study. Indeed, for the three sizes of particles, the experimental data collapse well on a master curve corresponding to the LLD law at a sufficiently large capillary number of the suspension $Ca_\phi$. 

These results also show that the threshold capillary number for the liquid film only regime is more challenging to predict, as emphasized earlier. Indeed, for a volume fraction $ \phi=41\%$, even at the lowest withdrawal velocity $U$, the particles of diameter $2\,a=40\,\mu{\rm m}$ are entrained in the film. There is no liquid only regime in the range of velocity accessible with our experimental setup. The formation of large clusters in the meniscus leads to the entrainment of particles at smaller capillary number than expected. At a smaller volume fraction, $\phi=18.6\%$, although we observe a regime where only the fluid is entrained on the plate, the transition is shifted to lower capillary numbers. The additional viscous forces induced by the presence of a large number of particles in the clusters thus lead to the formation of a heterogeneous coating on the plate at smaller capillary number than expected. In summary, the transition (1) between the liquid only and the heterogeneous film is highly dependent on the particle size and the volume fraction of the suspension.

 \section{Conclusion}

In this paper, we investigate the entrainment of a film of suspension on a plate withdrawn from a bath of non-Brownian particles dispersed in a Newtonian liquid at moderate volume fraction and without evaporation. For the Newtonian liquid only, the experimental results are captured by the classic LLD law, $h=0.94\,\ell_c\,Ca^{2/3}$. We show that the addition of particles to the fluid modifies this behavior.

Depending on the particle size and the volume fraction of the suspension, three regimes are observed. At low capillary number, \textit{i.e.}, small withdrawal velocity, we observe that only the liquid coats the plate if the suspension is sufficiently diluted. At large number capillary, \textit{i.e.}, high withdrawal velocity, we observe several layers of particles coating the plate. In this regime, the average thickness of the liquid film can be described by the LLD law accounting for the effective viscosity of the suspension $ \eta (\phi) $ through the capillary number $ Ca_\phi = \eta (\phi) \, U / \gamma $. Between these two regimes, a more complex regime is observed where the particles are entrained and form a highly heterogeneous coating. In this complex regime, particles tend to assemble into clusters.

We rationalize our experimental observations with theoretical predictions. The models provide a first estimate of the capillary numbers corresponding to the transitions between the regimes. These transitions depend both on the size of the particle and the volume fraction. The thickness of the liquid film follows the LLD scaling in both the liquid only regime (low capillary number) and in the effective viscosity regime (large capillary number).

A better characterization of the single particle entrainment in the vicinity of the meniscus could help to refine the threshold values of the capillary number but requires future work. In the heterogeneous regime, the structure of the deposited layer is complex. Ongoing work aims at describing the density of particle entrained, the clusters size and the thickness of the liquid film between the particles to better understand the physical mechanisms that control the properties of the heterogeneous coating. Nevertheless, our results provide an estimate of the operating parameters to obtain a homogeneous surface coating for a suspension of non-Brownian particles at intermediate volume fraction. Such homogeneous coating could be especially important in material science to provide a uniform surface property of the substrate. Our results also illustrate that a film of suspension can either be modeled with a continuum approach through an effective viscosity or with a particle-scale model accounting for the local capillary effects. The appropriate approach depends on the length scale of the liquid film compared to the diameter of the particles and the volume fraction of the suspension. 

\section*{Acknowledgements}
The authors acknowledge the support from the French ANR (project ProLiFic ANR-16-CE30-0009) and Saint-Gobain. The work of AS and ED is partially supported by a CNRS PICS grant n$^{\rm o}$07242. We thank H. A. Stone, C. Colosqui and G. M. Homsy for helpful comments and suggestions.





{\small \bibliography{rsc} 
    \bibliographystyle{ieeetr}}

\end{document}